\documentclass[%
 reprint,
 amsmath,amssymb,
 aps,
]{revtex4-2}

\usepackage{graphicx}
\usepackage{dcolumn}
\usepackage{bm}

\begin{document}

\preprint{APS/123-QED}

\title{Linking diffusive fields to virtual waves as their propagative duals}

\author{P. Burgholzer}
 \email{peter.burgholzer@recendt.at}
 \affiliation{Research Center for Non Destructive Testing (RECENDT), 4040 Linz, Austria}
  

 \author{L. Gahleitner}
\affiliation{University of Applied Sciences Upper Austria, 4600 Wels, Austria}

 \author{G. Mayr}
\affiliation{University of Applied Sciences Upper Austria, 4600 Wels, Austria}

\date{\today}

\begin{abstract}
In non-destructive and biomedical imaging, spatial patterns inside a sample are imaged without destroying it. Therefore, propagating waves, including electromagnetic or ultrasonic signals, or even diffuse heat are generated or modified by these internal patterns and transmit this structural information to the sample surface. There, the signals can be detected, and an image of the internal structure can be reconstructed from the measured signals. The amount of information about the interior of the sample that can be obtained from the detected signals at the sample surface is significantly influenced by the propagation from the internal structure to the surface. In the real world, all signal propagation is more or less irreversible. The entropy generated during propagation corresponds to the loss of information.

In an idealized model, such propagating waves, called virtual waves, are described by the wave equation. They remain valid solutions of this equation when the direction of time is reversed, thus exhibiting reversibility and showing no entropy production and therefore no loss of information during propagation.

A Fredholm integral equation of the first kind relates the diffusion fields locally for each point in space to the virtual waves as their propagating duals. These virtual waves can be calculated from the measured diffusion field at each detection point by inverting this integral equation. In the past, this Fredholm integral was derived using so-called “thermal waves” for thermography. This was often confusing because thermal diffusion cannot be described by the wave equation. Here we have derived this relationship in real space without having to use Fourier frequency space.

We have used the locally computed virtual waves from the measured diffusive surface signals for image reconstruction using established time-of-flight methods from ultrasound or RADAR imaging. This improves the spatial resolution in thermography and compensates for the dispersion of quantum wave packets in atom probe tomography.

\end{abstract}

\maketitle

\section{\label{sec:Introduction}Introduction}
In imaging, waves propagate from the structure to be imaged to a detector, where their signal is used to reconstruct the image. As already shown by Ernst Abbe in 1873, the spatial resolution is diffraction-limited, approximately by the wavelength\cite{Abbe}. In acoustics, e.g. in photoacoustic imaging, the detectable wavelength is limited not only by technical limitations such as the acoustic detector bandwidth but also by the frequency-dependent acoustic attenuation during propagation from the internal structure to the detector on the sample surface\cite{DeanBen.2011}. In liquids, the acoustic attenuation coefficient typically increases with the square of the frequency\cite{Patch.2007}. This leads to a broadening of the wave packets due to dissipation, in this case even without dispersion, as the speed of sound remains constant for all frequencies in the case of the square dependence of the attenuation on the frequency. 

In photoacoustic imaging, a short-pulse laser is usually used to illuminate a semi-transparent sample, such as biomedical tissue \cite{Wang.2009}. The light, which is absorbed to different degrees locally, causes a sudden increase in temperature and generates an ultrasonic pressure signal through immediate expansion. The photoacoustic reconstruction consists of calculating the spatial initial pressure distribution from the ultrasonic signals measured at several detection points outside the imaged sample. This initial pressure is locally proportional to the absorbed optical energy. Using ideal acoustic detectors and without acoustic attenuation the inverse reconstruction problem could be solved exactly, e.g. by time reversal\cite{Burgholzer.2007}. Such a time-invariant signal is referred to as a virtual wave signal, as in reality acoustic attenuation always leads to a higher loss of the higher-frequency components during wave propagation. For almost 20 years, there have been several attempts to compensate for this attenuation, e.g. by La Riviere et al.\cite{LaRiviere.October23292005, LaRiviere.2006}, by Ammari\cite{Ammari.2012}, by Dean-Ben et al.\cite{DeanBen.2011}, or by Burgholzer et al.\cite{Burgholzer.2020}. In a time discretized version, where the pressure as a function of time is a vector, this is achieved by introducing a linear relationship between the measured pressure signal vector $\mathbf{p}_\text{meas}$ and the virtual wave signal $\mathbf{p}_\text{virt}$  at the same detector location ("local transformation"), which would be a solution of the wave equation:
\begin{equation}
    \mathbf{p}_\text{meas} = \mathbf{M} \, \mathbf{p}_\text{virt}.
    \label{Eq:linear_relation}
\end{equation}
The matrix $\mathbf{M}$ can be calculated analytically only in a few special cases, for example for a quadratic frequency dependent aoustic attenuation\cite{Burgholzer.2019, Burgholzer.2020}. The determination of $\mathbf{p}_\text{virt}$ from the measured signal $\mathbf{p}_\text{meas}$ in Eq. (\ref{Eq:linear_relation}) constitutes an ill-posed or ill-conditioned inverse problem. The direct inversion of this equation would cause severe noise amplification resulting in useless solutions. Therefore, regularization methods are used for the inversion. This is a consequence of the information loss due to acoustic attenuation of higher frequency components during pressure propagation to the surface. 

We could transfer the virtual wave concept described by Eq. (\ref{Eq:linear_relation}) from attenuated acoustic waves with usually rather small attenuation during propagation of one wavelength to thermal diffusion\cite{Burgholzer.2017,JAPtutorial,Burgholzer2022}. The so-called "thermal waves", which are the frequency components in Fourier frequency space, are attenuated significantly by a factor of $exp(-2\pi) \approx 0.0019$ within one wavelength. The concept of partly compensation of wave packet broadening seems to be useful for non-destructive and biomedical imaging in general. For example, applications of a transformation of the diffusive electromagnetic wave into a wave field were shown by Lee et al. \cite{Lee.1989, Lee.1993} and for geophysical inverse problems by Gershenson \cite{Gershenson.1997}, who stated for the first time, that the virtual wave concept is also applicable to the interpretation of time dependent heat flow \cite{Gershenson1993}. He mentioned that Eq. (\ref{Eq:linear_relation}) could be used for the localization of remote time dependent heat sources, such as active volcanoes in geophysics, but he did not use it for thermographic imaging by ultrasound reconstruction from the virtual wave signals. Heat diffusion enables the imaging of structures inside the sample using thermography in photothermal imaging\cite{Burgholzer2022}. The wavelength of the thermal wave determines the spatial resolution. Higher frequency components with smaller wavelengths enable better spatial resolution. The higher attenuation of thermal waves compared to acoustics, however, leads to strong blurring and low spatial resolution when imaging deeper structures\cite{Burgholzer.2017b}. A unifying framework for treating diverse diffusion-related periodic phenomena under the global mathematical label of diffusion-wave fields has been developed by Mandelis\cite{Mandelis.2001}, like "thermal waves", charge-carrier-density waves, diffuse-photon-density waves, but also modulated eddy currents, neutron waves, or harmonic mass-transport diffusion waves.

Here we introduce the virtual wave concept without using the so-called "thermal waves" following the derivation of Romanov in section 6.3 of his book already introduced in 1986 \cite{Romanov}. In the study of inverse problems for equations of parabolic or elliptic types, he proved the possibility of going over to the study of an equivalent problem for equations of a hyperbolic type. We will use this for thermographic tomography and to compensate for the dispersion of quantum wave packets.

For the propagation of quantum particles the temporal evolution of the wave function is described by the Schrödinger equation. The width of a Gaussian wave packet, describing the propagation of a free or homogeneously accelerated quantum particle, increases proportional to propagation time\cite{Messiah,Shukla2010}. The physical reason for this is Heisenberg's uncertainty principle: as the wave packet localizes the quantum particle, the momentum, which is proportional to the speed of the quantum particle, cannot take on a specific value, but is distributed around a mean value. The corresponding virtual wave uses the ordinary wave equation instead of the Schrödinger equation and therefore exhibits no dispersion. Similar to acoustic damping and thermal diffusion, the broadening of the quantum packet due to dissipation can be compensated for by calculating a virtual wave for each detector location.

\section{\label{sec:VWC}The virtual wave concept for diffusion}

In this section we relate the diffusive wave field to virtual waves as their propagative duals. The resulting relation is the same one we got from thermal wave theory using the Fourier frequency space\cite{Burgholzer.2017}. The heat diffusion equation is a parabolic partial differential equation\cite{Carslaw.19591986printing}:

\begin{eqnarray}
       \frac{\partial }{\partial t} T(\mathbf{r},t) = \alpha \nabla^2 T(\mathbf{r},t),
       && \text {with initial condition} \nonumber \\
       T(\mathbf{r},t=0)=T_0(\mathbf{r})
        \label{Eq:HeatDiffusion}
\end{eqnarray}
where $T(\mathbf{r},t)$ is the temperature as a function of space and time, $\nabla^2$ is the Laplacian (second derivative in space), and $\alpha$ is the thermal diffusivity which is assumed to be homogeneous in the sample. This equation describes the temperature evaluation from heat diffusion for a time $t>0$ with initial temperature $T_0(\mathbf{r})$. The corresponding virtual wave $T_{virt}(\mathbf{r},t)$ as its propagative dual solves the wave equation of a hyperbolic type and has the same initial temperature. Since it is time reversible, it is even in $t$ and its first time derivative at $t=0$ must be zero:
\begin{eqnarray}
       \frac{\partial^2 }{\partial t^2} T_{virt}(\mathbf{r},t) = c^2 \nabla^2 T_{virt}(\mathbf{r},t) 
        && \text {with initial conditions} \nonumber \\
        T_{virt}(\mathbf{r},t=0)=T_0(\mathbf{r})
        &&  \nonumber \\
        \frac{\partial }{\partial t} T_{virt}(\mathbf{r},t=0) = 0,
        \label{Eq:VirtualWaveEquation}
\end{eqnarray}
where the propagation speed $c$ of the virtual waves can be chosen arbitrarily.

The measured surface temperature and the virtual wave are connected by a local transformation, i.e. for the same location $\mathbf{r}$:
\begin{eqnarray}
    T(\mathbf{r},t) &&= \int_{- \infty}^\infty T_{virt}(\mathbf{r},t') K(t,t')\text{d}t', \nonumber \\
    && \text{with} \nonumber \\
    K(t,t') &&\equiv \frac{c}{\sqrt{\pi \alpha t}} \exp{\left( - \frac{c^2 t'^2}{4 \alpha t} \right)} \phantom{X} \text{for} \phantom{X} t>0.
    \label{Eq:Temperature_virtual_wave}
\end{eqnarray}
This can be proven according to Romanov\cite{Romanov} by using:
\begin{eqnarray}
    \frac{\partial}{\partial t} K(t,t')= &&(-\frac{1}{2t}+ \frac{c^2 t'^2}{4 \alpha    t^2})K(t,t') \\ \nonumber
    \frac{\partial}{\partial t'} K(t,t')= &&-\frac{c^2 t'}{2 \alpha t}K(t,t') \\ \nonumber
    \frac{\partial^2}{\partial t'^2} K(t,t')= &&(-\frac{c^2}{2 \alpha t}+ (\frac{c^2 t'}{2 \alpha t})^2)K(t,t') \\ = &&\frac{c^2}{\alpha} \frac{\partial}{\partial t} K(t,t') \nonumber
    \label{Eq:KDerivative}
\end{eqnarray}

Using this last relation for $K(t,t')$ we can derive from Eq. (\ref{Eq:Temperature_virtual_wave}):
\begin{equation}
       \frac{\partial}{\partial t} T(\mathbf{r},t) = \int_{- \infty}^\infty T_{virt}(\mathbf{r},t') \frac{\alpha}{c^2} \frac{\partial^2}{\partial t'^2} K(t,t')\text{d}t'
       \label{Eq:K2Derivative}
\end{equation}

Two times integration by parts, using that $K(t,t')$ and $T_{virt}(\mathbf{r},t')$ goes to zero for $t'= \pm \infty$ and from the wave equation  (\ref{Eq:VirtualWaveEquation}) for $T_{virt}$ one gets:

\begin{eqnarray}
       \frac{\partial}{\partial t} T(\mathbf{r},t) = \int_{- \infty}^\infty \frac{\alpha}{c^2} K(t,t') \frac{\partial^2}{\partial t'^2} T_{virt}(\mathbf{r},t') \text{d}t' \nonumber \\
       = \int_{- \infty}^\infty \alpha K(t,t') \nabla^2 T_{virt}(\mathbf{r},t') \text{d}t' \nonumber \\
       = \alpha \nabla^2 \int_{- \infty}^\infty  K(t,t')  T_{virt}(\mathbf{r},t') \text{d}t' \nonumber \\
       = \alpha \nabla^2 T(\mathbf{r},t) 
       \label{Eq:K3Derivative}
\end{eqnarray}
For the last equation the Fredholm integral equation Eq. (\ref{Eq:Temperature_virtual_wave})  was used and therefore the temperature defined by the relation in Eq. (\ref{Eq:Temperature_virtual_wave}) fulfills the heat diffusion equation Eq. (\ref{Eq:HeatDiffusion}). The Fredholm integral equation of the first kind can be inverted to calculate the propagative dual $T_{virt}$ from the measured diffusive field $T$.
        
\section{\label{sec:photothermal}Photothermal reconstruction using the virtual wave signal}
Here we will show how to invert the Fredholm integral equation, Eq. (\ref{Eq:Temperature_virtual_wave}), to calculate the virtual wave $T_{virt}(\mathbf{r}_{surface},t)$ as its propagative dual from the measured surface temperature $T(\mathbf{r}_{surface},t)$ in a first step. In a second step, any ultrasonic reconstruction method, such as back-projection, Synthetic Aperture Focusing Technique (SAFT), or time reversal reconstruction, can be used to reconstruct the initial temperature $T_0^{rec}(\mathbf{r})$ from $T_{virt}(\mathbf{r}_{surface},t)$, which is the inversion of Eq. (\ref{Eq:VirtualWaveEquation}). The direct inversion of the heat diffusion in Eq. (\ref{Eq:HeatDiffusion}) could only be solved for very special cases such as planar layers parallel to the sample surface as a one-dimensional inverse problem.

\begin{figure}
\includegraphics[trim={7.5cm 3.0cm 7.2cm 2.5cm},clip,width=0.5\textwidth]{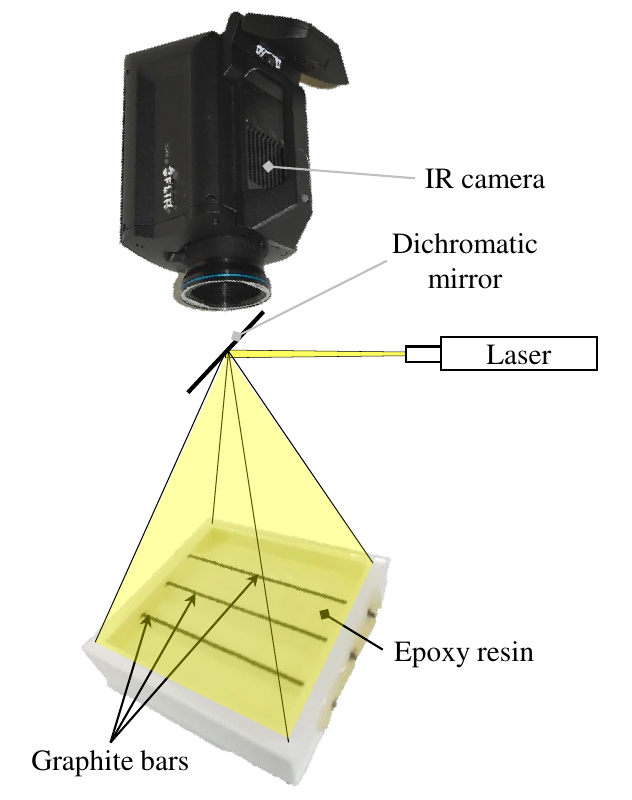}
    \caption{Sketched measurement set-up with three graphite bars embedded in epoxy resin at a depth of 1.6 mm, 2.6 mm, and 3.6mm, a laser for a short heating of the bars, and an infrared camera to measure the temporal evolution of the surface temperature. (Adapted from G. Thummerer et al., Photoacoustics 19, 100175, 2020; licensed under Creative Commons Attribution (CC BY) license.)}
    \label{fig:Bars}
\end{figure}
\begin{figure}
\includegraphics[trim={11.2cm 9.0cm 1.3cm 9.0cm},clip,width=0.5\textwidth]{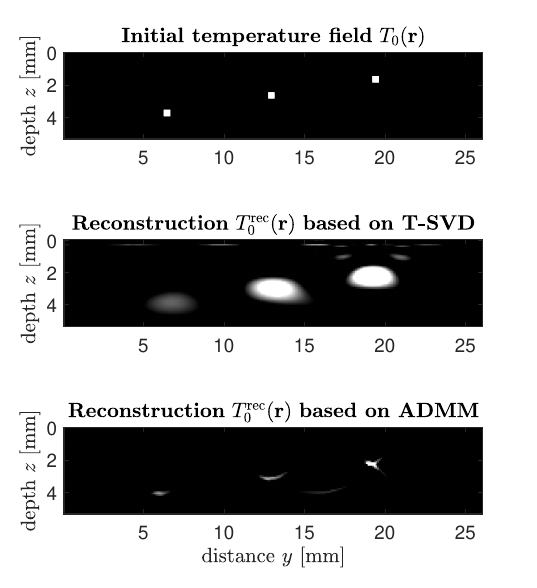}
    \caption{Comparison of the initial temperature distribution and the reconstructed initial temperature distributions using T-SVD or ADMM for calculating the virtual wave from the surface temperature measurement, followed by a subsequent reconstruction using SAFT. (Adapted from G. Thummerer et al., Photoacoustics 19, 100175, 2020; licensed under Creative Commons Attribution (CC BY) license.)}
    \label{fig:Barsrec}
\end{figure}

To demonstrate this photothermal reconstruction on a simple two-dimensional example, we have embedded three thin graphite bars parallel to the surface at depths of 1.6 mm, 2.6 mm, and 3.6 mm in an epoxy resin (Fig. \ref{fig:Bars}) \cite{Thummerer.2020b}. The epoxy is transparent to visible light but opaque to mid-infrared light. The graphite bars were heated by laser excitation (diode laser wavelength 938 nm, 250 W laser power with a pulse duration of 200 ms). The surface temperature after laser pulse excitation was measured with an infrared camera (106 Hz full frame mode and a noise equivalent temperature difference of 25 mK, cooled InSb sensor), which was sensitive in the spectral range of $3.0 - 5.1$ $\mu$m, where the epoxy resin is opaque.

The spatial resolution from the camera pixels was approx. 0.1 mm on the sample surface. Fig. \ref{fig:Barsrec} shows a cross section of the initial temperature field from the light absorbing bars, and the reconstruction using the truncated-singular value decomposition (T-SVD) method for calculating the virtual wave and SAFT for the ultrasound reconstruction algorithm. To calculate the virtual wave considering sparsity (the cross section of graphite rods are only small points) and positivity (heating always increases the temperature), the Alternating Direction Method of Multipliers (ADMM) as a nonlinear optimization algorithm was used\cite{Thummerer.2020b}. Sparsity is taken into account by minimizing in addition to a data fitting term also the L1-norm.  More precisely, the virtual wave is computed as a minimizer  of $\| \mathbf{K} \, T_{\text{virt}} - T\|^2 + \mathcal{R}(T_{\text{virt}}) $, where $\mathbf{K}$ is a discretization of the integral in Eq.  (\ref{Eq:Temperature_virtual_wave}) and $\mathcal{R}(.)$ is a suitable regularizer defined by the L1-norm. The use of this regularized virtual wave allows to implement efficient L1-minimization algorithms (e.g. ADMM) and leads to a much better resolution of the reconstructed temperature field \cite{Thummerer.2020b}.\par

Without additional information such as sparsity or positivity, spatial resolution degrades with depth\cite{Burgholzer2022}:
\begin{equation}
\delta_r(x) = \frac{\pi z}{\ln({SNR})}
    \label{Eq:delta_resolution_thermal}    
\end{equation}
\noindent
where $z$ is the depth of the graphite bars in epoxy of 1.6 mm, 2.6 mm, and 3.6 mm, ln is the natural logarithm, and $SNR$ is the signal-to-noise-ratio. The maximum temperature increase at the surface above the graphite bars is 1.3 K, 0.7 K, and 0.4 K. Using the noise equivalent temperature difference of 25 mK gives a $SNR$ of 52, 28, and 16, respectively. Averaging 300 cross-sections acquired parallel to the graphite bars in the $x$-direction, which are all measured for one excitation puls by the infrared camera, increases the $SNR$ by a factor of $\sqrt300$. Such direct averaging is not possible for three-dimensional structures. However, the second step in the reconstruction process, here the SAFT algorithm, is also an averaging over all camera pixels and therefore increases the SNR by the square root of the number of all camera pixels used for imaging. Thus, the SNR is increased by a factor of $\sqrt30000$. Using Eq. (\ref{Eq:delta_resolution_thermal}) gives a spatial resolution of 0.6 mm, 1 mm, and 1.4 mm for the reconstruction of the cross-section of the fibers in the axial direction (depth $z$) perpendicular to the surface. The lateral resolution (parallel to the sample surface in the $y$ direction) is a factor of approximately 2.7 worse than the axial resolution, which gives the elliptical shaped reconstructions for $T_0^{rec}(\mathbf{r})$ based on T-SVD in Fig. \ref{fig:Barsrec}\cite{Burgholzer2022}.

Where does this resolution limit come from? It is a result of entropy production or information loss during the diffusion process. Information is "transported" to the environment by entanglement and then lost by decoherence\cite{Burgholzer2022}. The probabilities that describe the uncertainties in the diffusion process turn out to come from quantum mechanics, as described in a Physics Today review by Katie Robertson\cite{PhysicsToday}. Three publications by different authors in 2006 show that quantum entanglement of the diffusive system with a large environment provide the probabilities\cite{PhysicsToday1,PhysicsToday2,PhysicsToday3}. So we wondered what would happen if we applied the virtual wave concept directly to a quantum wave packet describing the propagation of a free or homogeneously accelerated quantum particle. Its time evolution is described by the Schrödinger equation.

\section{\label{sec:quantum}Quantum wave packet propagation}
\begin{figure}
\includegraphics[width=\columnwidth]{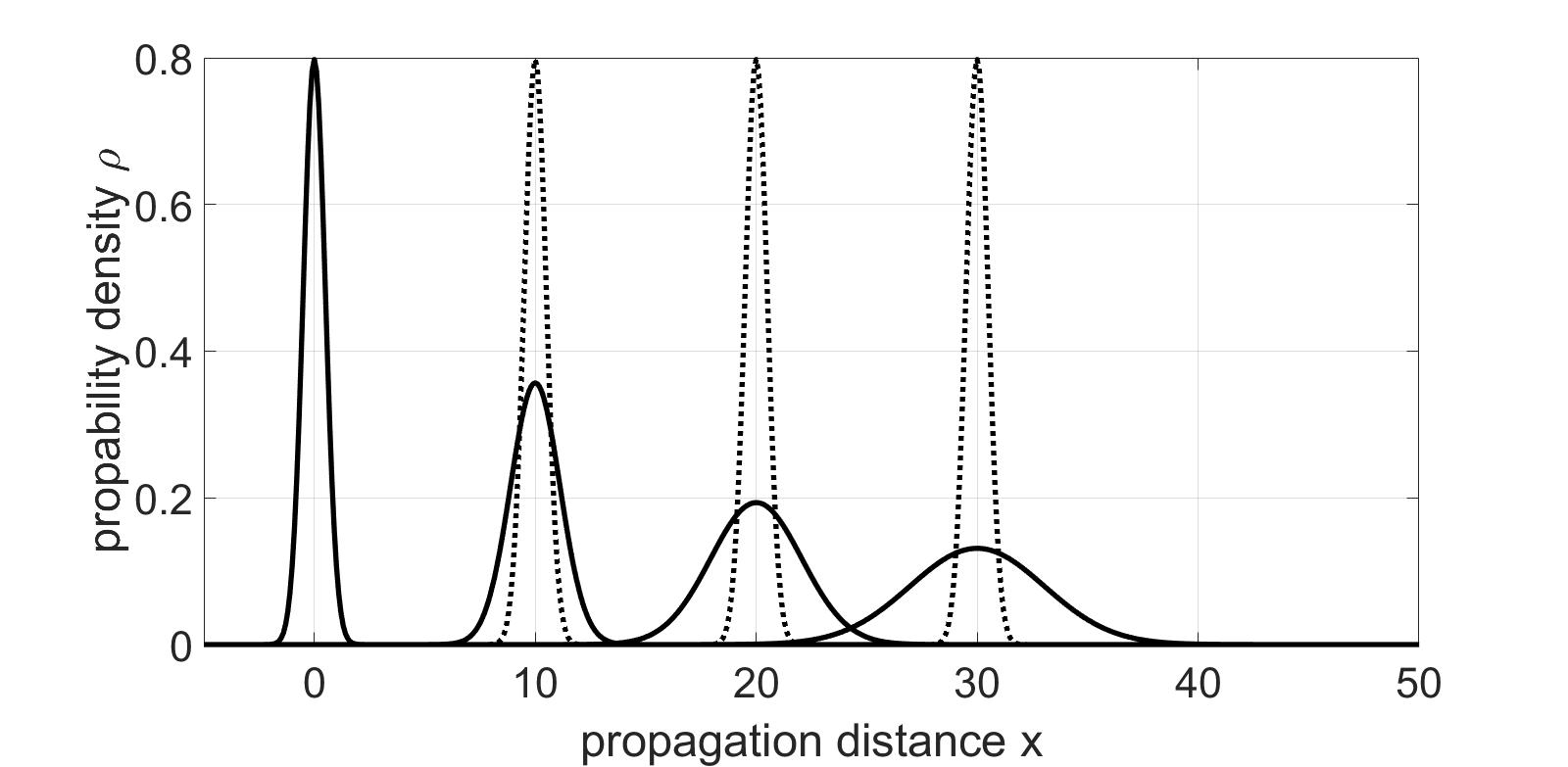}
    \caption{Evolution of the probability density $\rho=\psi^*\psi$ of a freely propagating Gaussian wave packet (solid line) shown for  for $m=\hbar=1$, $\sigma(0)=1/2$. The propagation speed for the corresponding virtual wave (dotted line) is set equal to the velocity of the wave packet $c=v=10$.
    The initial wave packet at $t=0$ has a width (square root of the variance) of $ 1/2$ and gets broader according to Eq. (\ref{Eq:width_wave_packet}), as shown for the times $t=1,2,3$.}
    \label{fig:wavepacket_density}
\end{figure}
Using the arguments from above, a propagating one-dimensional quantum wave packet having a mass $m$ and velocity 
$v$ without any interaction to the environment should suffer no information loss, as it stays a pure quantum state all the time. Nevertheless, the width as the standard deviation $\sigma$ of the probability density of a Gaussian wave packet increases proportionally to the propagation time\cite{Messiah}
\begin{equation}
\sigma(t)^2 = \sigma(0)^2+[\frac{\hbar t}{2 m \sigma(0)} ]^2,
    \label{Eq:width_wave_packet}    
\end{equation}
\noindent
as shown in Fig. \ref{fig:wavepacket_density} for $m=\hbar=1$, $\sigma(0)=1/2$ and $v=10$.
The probability density $\rho=\psi^*\psi$ is the square of the absolute value of the wave function $\psi(x,t)$. The time evolution of $\psi(x,t)$ is described by the free particle Schrödinger's equation,
\begin{eqnarray}
       \frac{\partial }{\partial t} \psi(x,t) = \frac{i\hbar}{2m} \nabla^2 \psi(x,t),
       && \text {with initial condition} \nonumber \\
       \psi(x,t=0)=\psi_0(x).
        \label{Eq:Schroedinger}
\end{eqnarray}
In comparison to the heat diffusion equation Eq. (\ref{Eq:HeatDiffusion}) the free particle Schrödinger equation for the wave function $\psi$ looks the same if the thermal diffusivity $\alpha$ is replaced by $i\hbar / (2m)$. Therefore, we can calculate the corresponding virtual wave by the local transformation (Fig. \ref{fig:wavepacket_density}):
\begin{eqnarray}
    \psi(x,t) &&= \int_{- \infty}^\infty \psi_\text{virt}(x,t') K_\psi(t,t')\text{d}t', \nonumber \\
    && \text{with} \nonumber \\
    K_\psi(t,t') &&\equiv c\sqrt{\frac{2m}{i \pi \hbar t}} \exp{\left(\frac{i m c^2 t'^2}{2 \hbar t} \right)} \phantom{X} \text{for} \phantom{X} t>0.
    \label{Eq:Quantum_wave}
\end{eqnarray}
Due to the "complex diffusivity" in comparison to $K$ in Eq. (\ref{Eq:Temperature_virtual_wave}) the $K_\psi$ in this equation shows no exponential decay but an oscillatory behavior. For discrete time steps the matrix $K_\psi$ can be inverted. For solving the inverse problem to get $\psi_\text{virt}$ from $\psi$ in Eq. (\ref{Eq:Quantum_wave}) no regularization is necessary because no information is lost during propagation. This seems to contradict the increasing width $\sigma$ of the wave packet shown for the solid line in Fig. \ref{fig:wavepacket_density} and described by Eq. (\ref{Eq:width_wave_packet}), because less resolution would result in reduced information. 

\begin{figure}
\includegraphics[width=\columnwidth]{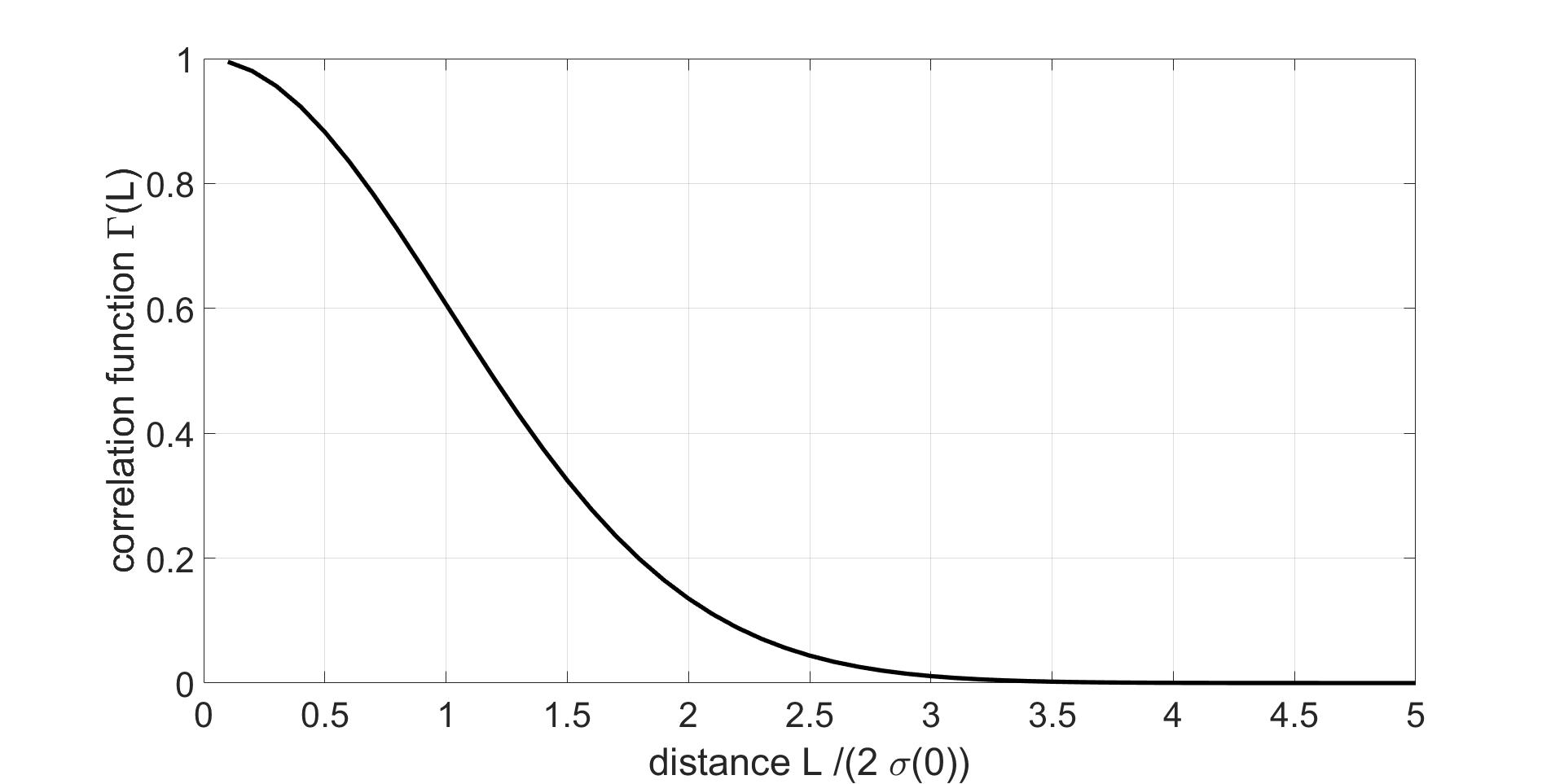}
    \caption{Correlation function of the wave packets from Fig. \ref{fig:wavepacket_density} at different times. It does not change with time.}
    \label{fig:quantum_correlation}
\end{figure}

The reason is that for the complex wave function $\psi(x,t)$ the longitudinal coherence 
\begin{equation}
 \Gamma(L) = \int_{- \infty}^\infty \psi^*(x,t) \psi(x+L,t)\text{d}x=\exp{(-{\frac{L^2}{8 \sigma(0)^2}})}
 \nonumber
    \label{Eq:wavepacket_correlation}    
\end{equation}
stays constant in time, which was shown e.g. by Klein et al.\cite{Klein} in neutron interferometry. In Fig. \ref{fig:quantum_correlation} the correlation is shown for the wave packets from Fig. \ref{fig:wavepacket_density} at different times and it does not change with time. The correlation length stays constant at $\overline{\mbox{L}}_\psi=8\sigma(0)^2$. 

For one-dimensional heat diffusion this is not the case. For a Gaussian temperature pulse, which is a solution of the heat diffusion equation (\ref{Eq:HeatDiffusion}), we get in the Fourier space
\begin{equation}
      T(x,t) = \frac{E}{2 \pi}\int_{- \infty}^\infty e^{i k x} e^{-k^2 \alpha t}\text{d}k = \frac{E}{2 \sqrt{\pi \alpha t}} e^{-\frac{x^2}{4 \alpha t}},
\nonumber
\end{equation}
with E is proportional to the energy of the pulse (unit [E]=Km) and its correlation function is
\begin{eqnarray}
 \Gamma(L) = \frac{1}{E^2}\int_{- \infty}^\infty T^*(x,t) T(x+L,t)\text{d}x
  \nonumber \\
 =\frac{1}{2 \pi}\int_{- \infty}^\infty e^{i k L} e^{-2k^2 \alpha t}\text{d}k 
   \nonumber \\= \frac{1}{2 \sqrt{2 \pi \alpha t}}\exp{(-{\frac{L^2}{8 \alpha t}})}, \nonumber
    \label{Eq:temperature_correlation}    
\end{eqnarray}
where Parseval's theorem is applied\cite{Klein}. The mean correlation length is
\begin{equation}
      \overline{\mbox{L}} = \int_{-\infty}^\infty |L| \Gamma(L)\text{d}L = 2\sqrt{\frac{2\alpha t}{\pi}} ,
      \nonumber
\end{equation}
which is not constant in time as for the quantum correlation, but increases with the square root of time. This is typical for diffusion processes and connected to the entropy production causing a loss in information and resolution.

Unlike an interferometer, a point-like detector has no "access" to the wave function but to the probability density. For the time evaluation of a free wave packet having a constant velocity $v$ or even with uniform acceleration $a$ another diffusion-like equation has been derived\cite{Shukla2010}:
\begin{eqnarray}
       \frac{\partial }{\partial t} \rho(x+vt+\frac{a}{2}t^2,t) = \frac{\hbar^2t}{4m^2\sigma(0)^2} \nabla^2 \rho(x+vt+\frac{a}{2}t^2,t),\nonumber\\
       \text {with initial condition} \nonumber \\
       \rho(x,t=0)=\rho_0(x).\nonumber
        \label{Eq:PDensity}
\end{eqnarray}
This results in the same solutions for the probability density as shown in Fig. \ref{fig:wavepacket_density} and the wave packet gets broader according to Eq. (\ref{Eq:width_wave_packet}). Now in comparison to the heat diffusion equation Eq. (\ref{Eq:HeatDiffusion}) the equation for the probability density $\rho$ looks the same if the thermal diffusivity $\alpha$ is replaced by the time dependent diffusion coefficient $D=\frac{\hbar^2t}{4m^2\sigma(0)^2}$. Therefore, we can calculate the corresponding virtual wave now directly from the probability density by inverting the local transformation:
\begin{eqnarray}
    \rho(x,t) &&= \int_{- \infty}^\infty \rho_\text{virt}(x,t') K_\rho(t,t')\text{d}t', \nonumber \\
    && \text{with} \nonumber \\
    K_\rho(t,t') &&\equiv \frac{c}{\sqrt{2\pi D t}} \exp{\left(-\frac{ c^2 t'^2}{2D t} \right)} \phantom{X} \text{for} \phantom{X} t>0.
    \label{Eq:Density_wave}
\end{eqnarray}
The probability density $\rho$ is locally connected to its virtual wave $\rho_\text{virt}$ and this can be derived similar to Eq. (\ref{Eq:Temperature_virtual_wave}). For the partial derivatives of $K_\rho$ it can be shown similar to Eq. (5):
\begin{equation}
      \frac{\partial^2}{\partial t'^2} K_\rho(t,t')=\frac{c^2}{D} \frac{\partial}{\partial t} K_\rho(t,t').
      \nonumber
\end{equation}
Following the analog procedure from Eq. (\ref{Eq:K2Derivative}) and (\ref{Eq:K3Derivative}) we proved Eq. (\ref{Eq:Density_wave}) by using the wave equation and $\rho_{virt}(x,t)=\rho_0(x-ct)$. Now the correlation function gets time dependent as in the case of heat diffusion, and the correlation length is not constant in time, but $\overline{\mbox{L}}_\rho=\frac{2}{\sqrt{\pi}}\sigma(t)$.

\section{\label{sec:apt}Information loss for quantum wave packets}
In comparison to the kernel $K_\psi$ for calculating the wave function using Eq. (\ref{Eq:Quantum_wave}), the kernel $K_\rho$ for calculating the probability density using Eq. (\ref{Eq:Density_wave}) shows no oscillatory behavior, but an exponential decay as the $K$ in Eq. (\ref{Eq:Temperature_virtual_wave}). Therefore, calculating the virtual wave from the measured probability density is again an ill-posed inverse problem which indicates an information loss compared to the wave function. 

But when does this information loss happen? For heat diffusion this is during the diffusion process. For the quantum wave packet this cannot occur during propagation, as this information loss would not happen if we use an interferometric measurement. The information loss occurs when the wave packet is detected by a point-like detector, as it happens e.g. in time-of-flight mass spectrometry or in atom probe tomography (APT)\cite{Gault2021}. Due to the interaction of the propagating wave packet with the detector all the information about velocity which could be seen in wavenumber-space is "transported" to the detector and its environment and then lost by decoherence. As this decoherence process with a macroscopic detector is short compared to the detection process this was described in earlier quantum mechanics as the "collapse" of the wave function during measurement\cite{Mueller}. The information loss due to detection is equal to the entropy production for a Gaussion wave packet $\Delta H(t)=ln(\sigma(t)/\sigma(0))$. Similar to heat diffusion the virtual quantum density in Eq. (\ref{Eq:Density_wave}) can be calculated from the measured quantum density by T-SVD or by taking the sparsity into account and using the ADMM for regularization, to get similar reconstructions as in Fig. \ref{fig:Barsrec} for thermal reconstruction.

In APT the time-of-flight mass spectrometry is used to figure out the mass-to-charge ratio of an ion by measuring its time-of-flight\cite{Gault2021}. The ions are accelerated by an electric field of known strength. Heavier ions with the same charge reach lower velocities. The speed of the ion depends on the ratio between its mass and charge. The time it takes for the ion to reach the detector is measured at a known distance. This time depends on the speed of the ion, so it's a way to measure its mass-to-charge ratio. With this ratio and the known experimental parameters, the ion can be identified.

\begin{figure}
\includegraphics[width=\columnwidth]{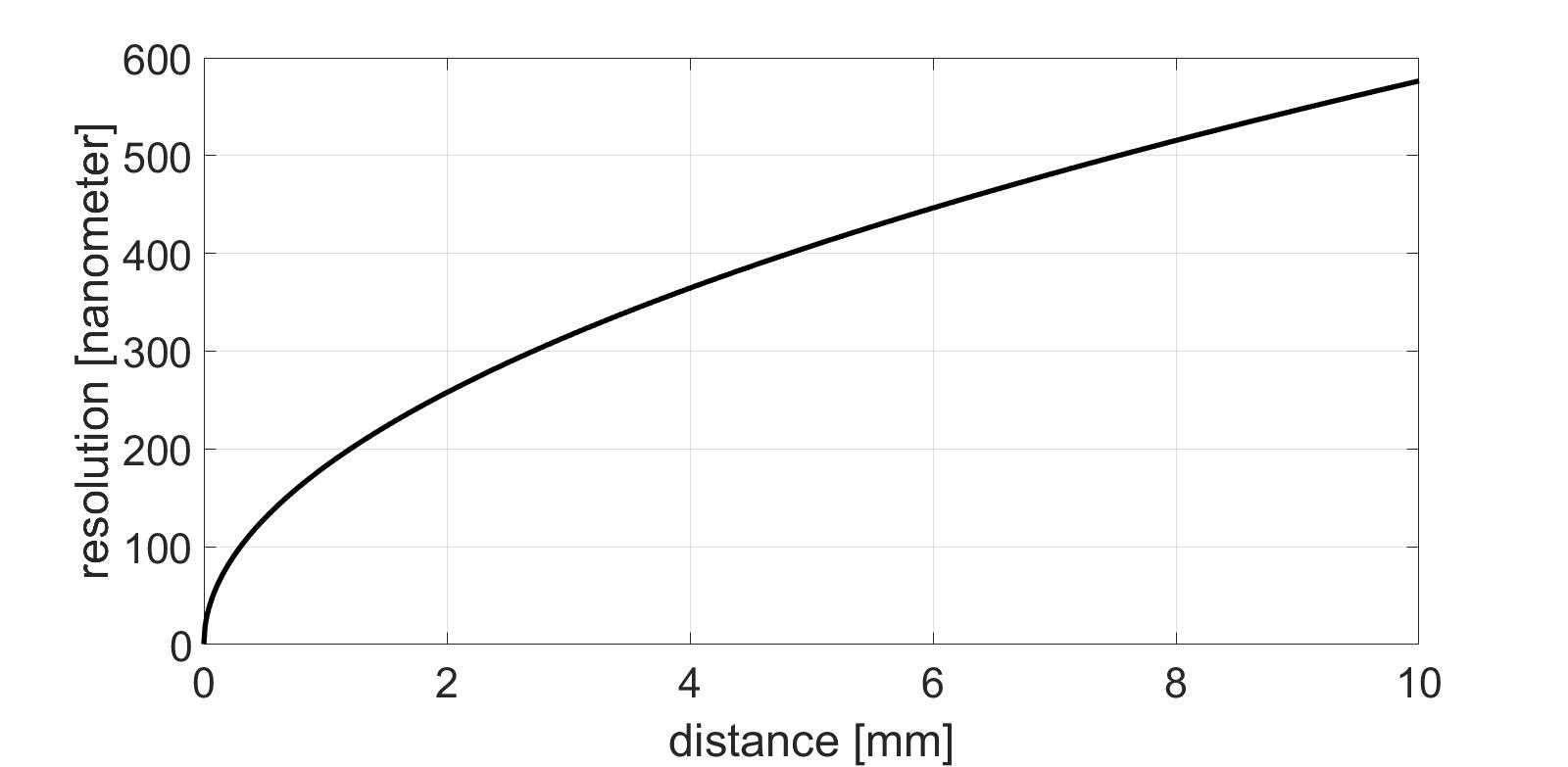}
    \caption{Spatial width and resolution of a wave packet defined as two times the standard deviation of $^{56}Fe^{2+}$ ions is shown as a function of the distance, calculated from Eq. (\ref{Eq:width_wave_packet}) and using an initial width of 0.3 nm from the distance of the Fe atoms. The accelerating electric voltage is 10 kV at a distance of 10 mm.}
    \label{fig:quantum_resolution}
\end{figure}

As an example in Fig. \ref{fig:quantum_resolution} the width and therefore also the spatial resolution of a wave packet of $^{56}Fe^{2+}$ ions are shown as a function of the distance. The accelerating electric voltage is 10 kV at a distance of 10 mm. The $^{56}Fe^{2+}$ ions were the most dominant Fe ionic species and showed more than 10000 counts for a hematite sample in APT\cite{Taylor}.

\section{\label{sec:conclusion}Conclusions and Outlook}
Fighting against entropy production and enhancing the signal-to-noise ratio (SNR) for measurements are closely connected goals, as described by the fluctuation-dissipation theorem\cite{JAPtutorial}. This is essential for successful reconstruction of diffusive fields in non-destructive and biomedical imaging. One way to accomplish this is by calculating the virtual wave as its propagative dual. The virtual wave is the solution of the wave equation that has the same initial conditions as the diffusive wave field but is reversible. Therefore, it is an idealized concept, as all wave propagation and diffusion processes in the real world are irreversible to a certain extent.

The virtual wave relates to the measurement results locally at each detection point via a Fredholm integral equation of the first kind. This equation is shown in Eq. (\ref{Eq:Temperature_virtual_wave}) for thermal diffusion and in Eq. (\ref{Eq:Density_wave}) for quantum wave packets measured in mass spectrometry or atomic probe tomography (APT). Calculating the virtual wave from the measurement results for a diffusive field is an ill-posed or ill-conditioned inverse problem that requires regularization. We used either truncated singular value decomposition (T-SVD) or implemented sparsity as additional information using the  Alternating Direction Method of Multipliers (ADMM), as shown in Fig. 2 for thermographic reconstruction. Even without considering additional information, such as sparsity, the virtual wave concept (VWC) allows one to perform a type of "averaging" over all detector pixels, enhancing the SNR.

Similar to the regularization methods used for thermographic imaging in Fig. 2, we propose using these methods for imaging with quantum wave packets in APT for future work.

\begin{acknowledgments}
This work was supported by the Austrian Research Funding Association (FFG) under Grant [905121]. This work is co-financed by the project “HyperMAT” by the federal government of Upper Austria and the European Regional Development Fund (EFRE) in the framework of the EU-program IBW/EFRE and JTF 2021-2027.
\end{acknowledgments}

\section*{DATA AVAILABILITY}
The data that support the findings of this study are available from the corresponding author upon reasonable request.

\section*{References}
\bibliographystyle{apsrev4-1}
\bibliography{JAP2022PB}

\end{document}